\documentstyle{camera}

\newcommand{\AmS}{{\protect\the\textfont2
  A\kern-.1667em\lower.5ex\hbox{M}\kern-.125emS}}

\newcommand{\beq}{\begin{equation}}
\newcommand{\eeq}{\end{equation}}
\newcommand{\bea}{\begin{eqnarray}}
\newcommand{\eea}{\end{eqnarray}}

\def\dm2{\Delta m^2}
\def\sq2{sin^2(2\Theta)}

\newcommand{\eg}{{\it e.g., }} 
\newcommand{\ie}{{\it i.e.}} 
\newcommand{\etal}{{\it et al.}}

\def \SAIT #1 #2 {{\em Mem.\ Soc.\ Astron.\ It.\/} {\bf #1}, #2}
\def \MESS #1 #2 {{\em The Messenger\/} {\bf #1}, #2}
\def \ASTRNACH #1 #2 {{\em Astron. Nach.\/} {\bf #1}, #2}
\def \AAP #1 #2 {{\em Astron. Astrophys.\/} {\bf #1}, #2}
\def \AAL #1 #2 {{\em Astron. Astrophys. Lett.\/} {\bf #1}, L#2}
\def \AAR #1 #2 {{\em Astron. Astrophys. Rev.\/} {\bf #1}, #2}
\def \AAS #1 #2 {{\em Astron. Astrophys. Suppl. Ser.\/} {\bf #1}, #2}
\def \AJ #1 #2 {{\em Astron. J.\/} {\bf #1}, #2}
\def \ANNREV #1 #2 {{\em Ann. Rev. Astron. Astrophys.\/} {\bf #1}, #2}
\def \APJ #1 #2 {{\em Astrophys. J.\/} {\bf #1}, #2}
\def \APJL #1 #2 {{\em Astrophys.. J. Lett.\/} {\bf #1}, L#2}
\def \APJS #1 #2 {{\em Astrophys. J. Suppl.\/} {\bf #1}, #2}
\def \APSS #1 #2 {{\em Astrophys. Space Sci.\/} {\bf #1}, #2}
\def \ASR #1 #2 {{\em Adv. Space Res.\/} {\bf #1}, #2}
\def \BAIC #1 #2 {{\em Bull. Astron. Inst. Czechosl.\/} {\bf #1}, #2}
\def \JSQRT #1 #2 {{\em J. Quant. Spectrosc. Radiat. Transfer\/} {\bf #1}, #2}
\def \MN #1 #2 {{\em Mon. Not. R. Astr. Soc.\/} {\bf #1}, #2}
\def \MEM #1 #2 {{\em Mem. R. Astr. Soc.\/} {\bf #1}, #2}
\def \PLR #1 #2 {{\em Phys. Lett. Rev.\/} {\bf #1}, #2}
\def \PASJ #1 #2 {{\em Publ. Astron. Soc. Japan\/} {\bf #1}, #2}
\def \PASP #1 #2 {{\em Publ. Astr. Soc. Pacific\/} {\bf #1}, #2}
\def \NAT #1 #2 {{\em Nature\/} {\bf #1}, #2}
\input epsf

\begin{document}

%
\title{HIGH REDSHIFT SUPERNOVAE: COSMOLOGICAL
IMPLICATIONS}

%
\author{NINO PANAGIA}

%
\organization{ESA/Space Telescope Science Institute, 3700 San Martin
Drive, Baltimore, MD 21218, USA; {\it panagia@stsci.edu}}

\maketitle

\begin{abstract}

We review the findings for the values of the cosmological parameters as
derived from high-redshift SNIa measurements. The most  recent results
confirm the picture of a non-empty inflationary Universe that is
consistent with a cosmological constant $\Omega_\Lambda\simeq0.7$. 
This  implies that the expansion of the  Universe is currently
accelerated by the action of  some mysterious dark energy.  We also
discuss the possibility and the consequences of the fact that SNIa may
not be {\it perfect} standard candles, in the sense of having properties
in the early Universe that are systematically different from those they
have at the present times.

\end{abstract}
\vspace{1.0cm}

\section{Introduction}

In the early part of this century, the pioneering work by Hubble, who
first observed that galaxies were receding from each other and
concluded that the Universe had to be expanding, led to a revolution
in our fundamental understanding of the Universe. In the 1960's, the
discovery of the Cosmic Background Radiation provided a physical
foundation for the expanding Universe.  In the 1970's and 1980's, the
emerging model, i.e., the hot Big-Bang became well established. 

Currently, a large number of fundamental physical and astrophysical
observations and theories are providing the foundation and tests for
the hot Big-Bang model. In order to compare models with observations
meaningfully, one needs accurate and reliable values of important
cosmological parameters. With this motivation, special efforts have
been made in the course of the years in this direction and recently
enormous progress has been made to measure H$_0$ and the deceleration
parameters with small errors. 

To carry-out these measurements, one needs distance indicators whose
intrinsic brightness is sufficiently well known and which are bright
enough to be seen at cosmologically significant recession velocities
i.e., well away  (far enough) from any local velocity perturbation.
Since supernovae are very bright objects and are relatively common (in
an ``average cluster of galaxies" with, say, $10^{13}$~M$_\odot$ in
stars, one may expect several SN explosions to occur per year), they
constitute prime candidates to probe distances to galaxies.

\section{Supernovae of Type Ia and the Expansion of the 
Universe} 

Supernovae of a particular type, denoted as Type Ia supernovae (SNIa)
and characterized by the  absence of hydrogen in their envelopes, are
perhaps the best ``standard candles": they are so  similar to each other
that their brightness provides a dependable gauge of their distance, and
so  bright that they are visible billions of light years away.  However,
knowing that they are all alike  is not enough, one needs also to
determine the exact value of their luminosity before using them  as
proper standard candles. That SNIa could be used as standard candles has
been proposed for  many years(\eg Kowal 1968). However, most of  the
progress in this field has occurred over the last decade.

Extensive ground-based surveys have identified a large number of new
supernovae and  characterized their global properties in a
statistically meaningful way. At the same time, using  the Hubble Space
Telescope, an international team led by Allan Sandage, and including
Gustav  Tamman, Abi Saha, Duccio
Macchetto and the author,  has carried out an extensive program to
determine the absolute brightness of a selected sample of  supernovae.
This has allowed to place SNIa on an absolute scale and the expansion
rate of the  Universe, the so-called Hubble constant to be determined
with a precision of 10\%. The derived  value for the Hubble constant is
$H_0=59\pm6 ~km~s^{-1}Mpc^{-1}$ (Saha \etal~2001, Tammann \etal~2002). 
This implies possible ages of the Universe in  the range 11-17 billion
years, depending of the acceleration/deceleration history of the
Universe  itself. 

\begin{figure} 
\centerline{\epsfxsize=10.cm\epsfbox{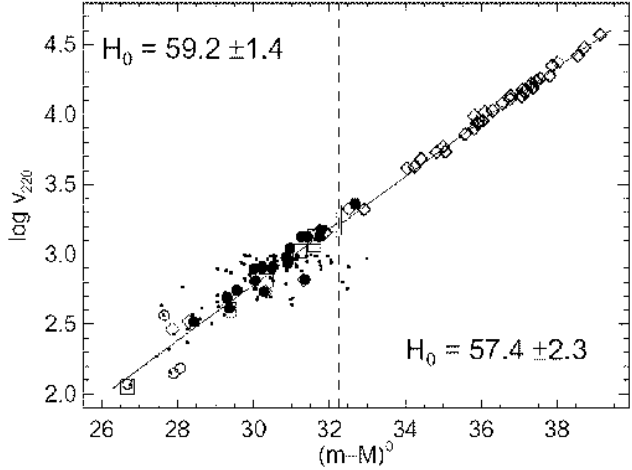}}
\caption{ The Hubble diagram based on: left hand panel - Cepheids (large dots), TF
distances (small dots), and galaxy groups/clusters (squares); right hand
panel - SNIa calibrated distances (Tammann \etal~2002).}
\end{figure}

However, even SNe\,Ia are not perfect standard candles. Their
absolute magnitudes correlate with second parameters, like decline rate
$\Delta m_{15}$, light curve shape, SN color, spectral features
(possibly correlated with temperature), and Hubble type or color of the
parent galaxy. Starting with Pskovskii (1977) and Phillips (1993), there
exists a rich literature on the subject (for a review, see Parodi \etal~
2000). Different empirical relations have been derived to homogenize the
SN data, but there is no clear picture of the underlying physics (see
\eg H\"oflich \& Khokhlov 1996).  The second-parameter corrections are a
serious problem because the nine calibrating SNe\,Ia have slower decline
rates, are bluer, and lie in later-type galaxies than the distant
SNe\,Ia.  Consequently the resulting value of $H_0$ varies depending on
which correction is adopted. The problem is accentuated by the fact that
some authors exclude some of the calibrators on grounds of their age or
{\it alleged} quality of their photometry, and/or include calibrators
without direct Cepheid distances. As a result current determinations of
$H_0$ based on SNe~Ia data vary between 55 (Schaefer 1998) and 71
(Freedman \etal~2001). According to a recent discussion by Panagia
(2003a), the most probable value appears to be  $H_0=63\pm6
~km~s^{-1}Mpc^{-1}$.

\section{Is the Universe accelerated?}

Once the expansion rate is determined, the next step is to determine
its variation in the course of  the evolution of the Universe. It is
interesting to note that, since one wants to measure the  variation of
the expansion rate, it is only necessary to verify that SNIa are
indeed standard  candles, that is that they have the same properties at
all redshifts (or, equivalently, at all look- back times), with no need
to determine their absolute luminosities. In other words, one could 
measure the deceleration parameters without measuring $H_0$, and
viceversa. The acceleration of  the Universe can be determined using
SNIa as standard candles, if these are observed at suitably  large
distances in order to reveal a measurable deviation from a constant
expansion.

Pioneering work by Danish astronomers N{\o}rgaard-Nielsen and
collaborators in the late 1980's  led to the discovery of one SNIa at
redshift 0.31 as the result of a multi-year observational effort 
(N{\o}rgaard-Nielsen \etal~1989).  It was only when Perlmutter's
Supernova Cosmology Project (SCP) took off that SNIa searches  at high
redshifts became an efficient reality. The SCP is an international
collaboration of  researchers from the United States, Sweden, France,
the United Kingdom, Chile, Japan, and  Spain. Thanks to their use of
large format detectors, a ``clever" observational strategy, and 
sophisticated image analysis techniques (Goobar \& Perlmutter 1995), in
1992 they discovered their first high redshift SNIa,  SN 1992bi at
$z=0.46$ (Perlmutter \etal~1995), followed in 1994 by 6 more at
$z>0.35$  (Perlmutter \etal~1997). Currently they discover about a 
dozen SNIa during each observing run, typically twice a year. 

Friendly competition promptly followed suit, when the High-Z Supernova
Search Team, another  international collaboration led by Brian Schmidt
(Australian National University) started their  systematic searches in
1995, essentially adopting Perlmutter's strategy and, consistently,
also  discovering a dozen supernova candidates per run. By mid-1998 the
SCP team had discovered  and studied about 78 SNIa (Perlmutter \etal~
1998, 1999), most of which in the redshift range 0.2--0.8, and the HZSS
team had  discovered about 32 SNIa (Riess \etal~1998). 

In the last few years, both groups have augmented their SNIa samples
and, for selected SNIa,  have complemented their ground-based
observations with much higher quality photometry obtained  with the {\it
HST--WFPC2\/}. The obvious improvement provided by the superior
angular  resolution of HST ~ is that contamination by the host galaxy
light is significantly reduced and  becomes an easily treatable
problem, thus providing higher photometric accuracy (see Figure  2).

\begin{figure} 
\centerline{\epsfxsize=12.cm\epsfbox{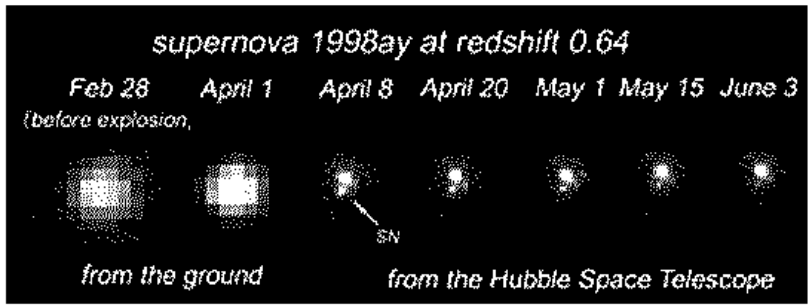}}
\caption{ Sequence of ground-based and HST-WFPC2 observations of SN
1998ay (SCP Team; see also Knop \etal~2003)}
\end{figure}

When SCP and HZSS researchers initially set out to measure the
expansion rate of the Universe,  they expected to find that distant
supernovae appeared brighter than their redshifts would  suggest,
indicating a slowing rate of expansion with time as the effect of
gravitational attraction  of all masses in the Universe. Instead they
found the opposite: at a given redshift, distant   supernovae were
dimmer than expected (see Figure 3). Expansion was accelerating!

The SNIa evidence for acceleration was tightened  up by the discovery  and
the study of SNIa at redshifts higher than 1. The most distant SNIa 
found so far is SN 1997ff. The supernova was discovered by Gilliland and
Phillips (1998; see also Gilliland, Nugent \&  Phillips 1999) in a
repeat  HST-WFPC2 observation of the Hubble Deep Field North (HDF-N) and
was serendipitously  monitored with HST-NICMOS throughout Thompson and
collaborators Guaranteed-Time- Observer campaign.  Analysing the
available HST data, Riess and collaborators (2001) determined the 
supernova type and redshift from the observed colors and their temporal
behavior, both of which  match a typical SNIa at redshift about 1.7.
This makes SN 1997ff the farthest supernova  observed to date. Fits to
observations of the SN provide constraints for the redshift-distance 
relation of SNIa that lend strong support to the currently preferred
cosmological model, with  about 30\% mass density and 70\% dark energy. 

The results obtained for additional high redshift SNIa (Knop \etal~2003,
Tonry \etal~2003)  provide strong support to the cosmological
interpretation (see Figure 3)  while excluding conclusively  grey dust
absorption and/or monotonically varying evolutionary effects, which
could explain the dimming (relative to the expections for an empty
Universe) of SNIa  up to $z\sim0.5$, but definitely not the  brightening
as found for higher z SNIa. 

\begin{figure} 
\centerline{\epsfxsize=11.cm\epsfbox{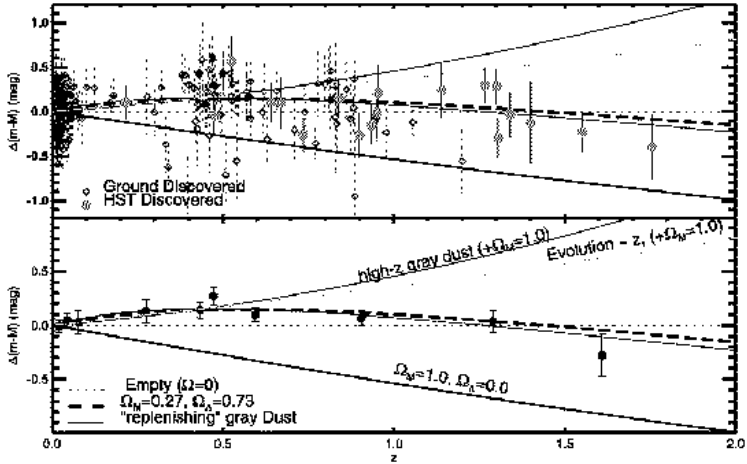}}
\caption{  SN Ia residual Hubble diagram comparing cosmological models
and models for astrophysical dimming. Top: SNIa from ground-based
discoveries in the top-quality sample are shown as diamonds;
HST-discovered SNIa are shown as filled symbols. Bottom: Weighted
averages in fixed redshift bins are given for illustrative purposes
only. Data and models are shown relative to an empty universe model
($\Omega = 0$). High redshift supernovae exclude grey dust and/or
monotonic evolutionary effects to explain the SNIa data and strongly
favor the cosmological interpretation with $\Omega_M\sim0.3$,
$\Omega_\Lambda\sim0.7$ (Riess \etal~2004).}
\end{figure}

Most recently Riess \etal~(2004b) reported the results of an HST search
of high z SNe. The search exploited the repeated observations (five
epochs at about 45 day intervals) of two fields of approximately
$10'\times15'$, centered on the Hubble Deep Field North (HDF-N) and the
Chandra Deep Field South, respectively, made by the  Great Observatories
Origins Deep Survey (GOODS) Hubble Space Telescope Treasury Program with
the Advanced Camera for Surveys (Giavalisco \etal~2004). Photometric
redshift measurements of the hosts combined with deep F606W, F775W, and
F850LP imaging were used to discriminate hydrogen-rich SNe II from SNe I
at $z>1$ on the basis of a marked UV deficit in the energy distributions
of SNIa (Panagia 2003b, Riess \etal~2004a).  Subsequent spectroscopy
of 11 GOODS SNIa obtained from the ground and with the grism on ACS
confirmed the reliability of the photometric screening. 

\begin{figure}[t!] 
\centerline{\epsfxsize=9.cm\epsfbox{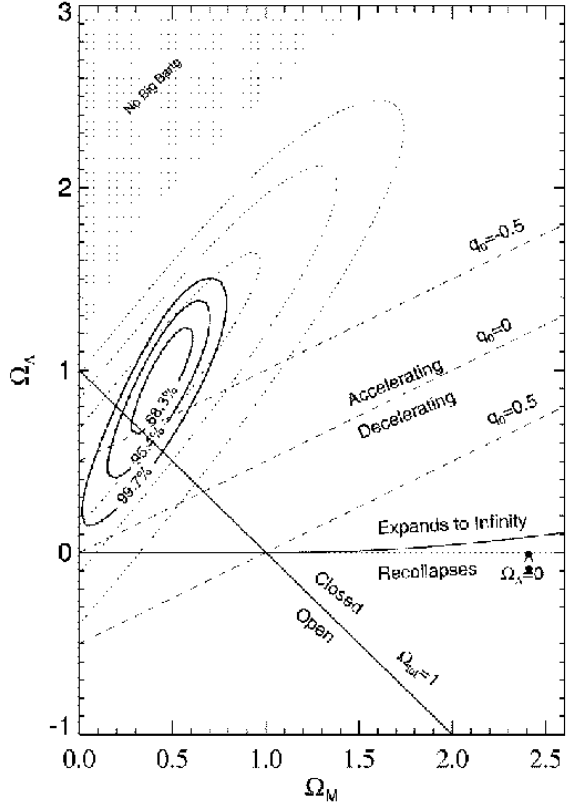}}
\caption{ Constraints to the matter density $\Omega_M$ and the vacuum  energy
density $\Omega_\Lambda$ as derived from the most recent SNIa observations by 
Riess et al. (2004).}
\end{figure}

In this way 16 Type Ia supernovae were discovered and used  to
provide the first statistically ``convincing" evidence for cosmic
deceleration that preceded the current epoch of cosmic acceleration.
These objects include 6 of the 7 highest redshift SNIa known, all at
$z>1.25$, and populate the Hubble diagram in unexplored territory. The
luminosity distances to these objects and to 170 previously reported SNe
Ia (Perlmutter \etal~1998, Riess \etal~1998, Perlmutter \etal~1999,
Tonry \etal~2003, Knop \etal~2003) have been determined using empirical
relations between light-curve shape and luminosity, based on local
Universe measurements assumed to apply to high redshift objects as
well.  A purely kinematic interpretation of the SN Ia sample provides
evidence for a transition  from deceleration to acceleration that is 
constrained to be at z=$0.46\pm0.13$. The data are consistent with the
cosmic concordance model of $\Omega_M\sim0.3$,$\Omega_\Lambda\sim0.7$ 
and are inconsistent with a simple model of monotonic evolution  with
redshift, or extinction by grey dust as an alternative to dark energy. 

For a flat Universe ($\Omega_tot=1$; de Bernardis \etal~2000, 2002,
Spergel  \etal~2003) with a cosmological constant, Riess \etal~(2004) 
measure $\Omega_M\sim0.29\pm0.04$ (equivalently, $\Omega_\Lambda=0.71$).
When combined with external  constraints imposed by  cosmic microwave
background and large-scale structure measurements (\eg Turner 2002), they
find $w=-1.02\pm0.16$ (and $w<-0.76$  at the 95\% confidence level) for
an assumed static equation of state of dark energy, $P=w\rho c^2$.These
constraints are consistent with the static nature and value of $w$
expected for a cosmological constant (i.e., $w_0=-1.0, ~dw/dz=0$) and
are inconsistent with a very rapid evolution of dark energy. 

Thus, it appears that the Universe would never come to an end, and more 
fundamentally that a large part of the Universe is made of  something we
know  nothing about -- the mysterious whatever-it-is that goes by the 
name of``dark energy" -- that  approximately 7 billion years ago
overcame gravity and started pushing the Universe to an  accelerated
expansion. 

The first attempt to explain the nature of dark energy was by invoking
Albert Einstein's notorious  ``cosmological constant," an extra term he
introduced in the equations of the theory of general  relativity early
in the 20th century under the mistaken impression, shared by astronomers
and  cosmologists of the time, that the Universe was static. The
cosmological constant, which Einstein  signified by the Greek letter
$\Lambda$, made it so.  Einstein happily abandoned the cosmological
constant when, in 1929, Edwin Hubble found that the  Universe was not
static but expanding. However, $\Lambda$ came back strong - albeit 70
years  later -  when supernova studies led to the discovery that
expansion was accelerating.

These results are rather unexpected and puzzling: for example, the fact
that the cosmological  constant value is comparable to the current mass
density (which decreases with time) would  place us at a ``special" time
in the evolution of the Universe. It is clear that the problem is far 
from solved, but can be solved: one needs to study more SNIa, over a
wider range of redshifts to  reduce the uncertainty region and to test
for the presence of possibly ``unseen" systematic effects,  e.g.,
evolution of the SNIa properties with redshift and/or metallicity.

Currently, there are two leading interpretations for the dark energy as
well as many more exotic  possibilities. It could be an energy
percolating from empty space as Einstein's theorized  ``cosmological
constant," an interpretation which predicts that dark energy is
unchanging and of a  prescribed strength. 

An alternative possibility is that dark energy is associated with a
changing  energy field dubbed ``quintessence" (see \eg  Caldwell, Dave,
\& Steinhardt 1998). This field would be causing the current
acceleration -- — a milder version of the inflationary  episode from
which the early Universe emerged. When astronomers first realized the
Universe  was accelerating, the conventional wisdom was that it would
expand forever. However, until we  better understand the nature of dark
energy and  its properties, other scenarios for the fate of the 
Universe are possible. If the repulsion from dark energy is or becomes
stronger than Einstein's prediction, the Universe  may be torn apart by
a future ``Big Rip" (see \eg Caldwell, Kamionkowski, \& Weinberg 2003),
during which the Universe expands so violently that  first the galaxies,
then the stars, then planets, and finally atoms come unglued in a
catastrophic  end of time. Currently this idea is very speculative, but
being pursued by theorists. At the other extreme, a variable dark energy
might fade away and then flip in force such that it  pulls the Universe
together rather then pushing it apart. This would lead to a ``Big
Crunch" (\eg Endean 1997) where  the Universe ultimately implodes.

\section{Is there any room for a doubt?}

As summarized by Riess \etal~(2004), the potential for luminosity evolution 
of corrected SN Ia distances
has been studied using a wide range of local host environments. No
dependence of the distance measures on the host morphology, mean stellar
age, radial distance from the center, dust content, or mean metallicity
has been seen (Riess \etal~ 1998; Perlmutter \etal~ 1999; Hamuy \etal~
2000). No differences in the inferred cosmology were seen by Sullivan
\etal~ (2003) for SNIa in early-type hosts or late-type hosts at high
redshifts. These studies limit morphology dependence of SN Ia distances
to the 5\% level. Detailed studies of distance-independent observables of
SNIa, such as their spectral energy distribution and temporal
progression, have also been employed as probes of evolution (see Riess
2000, Leibundgut 2001, and Perlmutter \& Schmidt 2004 for reviews).
The current consensus is that there is no evidence for evolution
with limits at or below the statistical constraints on the average
high-redshift apparent brightness of SNIa. The observed nominal
dispersion of high-redshift SNIa substantially limits the patchiness
of uncorrected extinction, and near-IR observations of a high-redshift
SN Ia demonstrate that a large opacity from grayish dust is unlikely
(Riess et al. 2001).

The case, therefore, would appear to be water-tight as long as {\it all}
SNIa events have one and the same origin.  However, this may not be
quite true.

\subsection{SNIa properties in the local Universe}

In a recent study, Mannucci \etal~(2005) have determined the rate of
supernovae (SNe)  of different types along the Hubble sequence as a
function of both the near-infrared luminosity and the stellar mass of
the parent galaxies.  They find that the rates of all SN types,
including Ia, Ib/c and II, show a sharp dependence on both the
morphology and the (B--K) colors of the parent galaxies and, therefore,
on the star formation activity.  In particular the SN Ia rate in late
type galaxies turns out to be  a factor $\sim$20 higher than in E/S0
galaxies. Similarly, the SN Ia rate in  galaxies bluer than B--K=2.6 is
about a factor of 30 greater than in galaxies with B--K$>$4.1.  These
findings are clear evidence that a significant fraction of Ia events in
late Spirals/Irregulars originates from a relatively young stellar
component.

An independent indication of different channels to produce SNIa
explosions is provided by the study of the frequency of SNIa events
occurred in elliptical galaxies.  An analysis of SNIa events in large
sample of early type galaxies (Della Valle \& Panagia 2003, Della Valle
\etal~2005) unambiguously shows that the rate of type Ia Supernovae
(SNe) in radio-loud galaxies is about 4 times higher than the rate
measured in radio-quiet galaxies. The actual value of the enhancement is
likely to be in the range $\sim 2-7$ (P$\sim 10^{-4}$). 

Discussing the possible causes of the SNIa rate enhancement in
radio-loud ellipticals, Della Valle \etal~(2005) conclude that this
phenomenon has the same common origin that determines these galaxies to
be strong radio sources, but that there is no causality link between the two
phenomena.  In particular, they argue that repeated episodes of
interaction and/or mergers of early type galaxies with dwarf companions,
on times-scale of $\sim 1$Gyr, are responsible for inducing both strong
radio activity in early type galaxies and to supply an adequate number
of ``fresh" SNIa progenitors to the old (Population II) stellar
population of ellipticals.

\subsection{Will two types of SNIa spoil the party?}

Considering the systematic diversity of SNIa in the local Universe, 
the use of one and the same kind of SN Ia as cosmological standard candles
may be questioned (for a more detailed discussion, see Mannucci, Della 
Valle \& Panagia 2005). 

From the study by Mannucci \etal~(2005), it appears that  in the local
Universe there are two  distinct populations of  progenitors of SNIa,
each characterized by very different delay times, \ie~characteristic
times between star formation and stellar explosion.  About half of the
SNIa originate from a relatively young stellar population, whose
progenitors explode after a short delay time (say, $<$100~Myrs) since
their formation({\it ``prompt"} SNIa),  and may be the result of the merging
of two degenerate stars in a binary system. The other half ({\it
``tardy"} SNIa) comes from  older stellar populations and their
progenitors explode with delay times of 2--4 Gyrs (\eg Strolger \etal~
2004, Gal-Yam \& Maoz 2004),  and possibly originate from a binary
system  composed of a white dwarf and an ordinary star.  It is easy to
realize that these two species of SN Ia will not be equally frequent at
all redshifts: the {\it prompt} exploders dominate at high redshifts
where the star formation is particularly active, whereas  the {\it
tardy} exploders become the dominant species in the  less distant
Universe where the overall star formation rate is rapidly declining. 

This would not be a problem if it was not that SNIa originating in
different environments appear to have peak luminosities appreciably
different from each other. In particular, in the local Universe, SNIa
occurring in late type galaxies are found to be  systematically brighter
by several tenths of a magnitude  than SNIa occurring in early type
galaxies (Filippenko 1989, Della Valle \& Panagia 1992, Hamuy \etal~1996,
Howell 2001).   Extending this trend to all redshifts, one would predict
the alarming effect that SNIa at redshifts less than 0.5-1 would be {\it
intrinsically} dimmer than SNIa occurring at higher redshifts, $z>1$,
which is just the behaviour  that is regarded as proof for
acceleration.  As a result, all quantitative conclusions about the
acceleration of the Universe would have to be drastically revised, if
not even reversed (\ie, there might {\it not} be any acceleration...).

Curiously enough, Sullivan \etal~ (2003), from a thorough analysis of
all samples available at the time,  found marginal evidence that at high
redshifts SNIa occurring in early type galaxies are  $0.14 \pm 0.09$
magnitudes brighter than the ones occurring in late type galaxies. They
attribute this difference to the effect of dust extinction that is non
negligible in late type galaxies whereas early type galaxies are
essentially dust-free.

The fact that a fair extrapolation of well established properties
observed for SNIa in the local Universe be at variance with what appears
to be the case at high redshifts is puzzling.  It could indicate that
there are unsuspected evolutionary effects that end up canceling each
other out, or that the statistics and/or the accuracy of the
measurements of SNIa events is not adequate to provide an unambiguous
answer.  In both cases, it appears that the problem of whether there is
an acceleration in the Universe, and if so what its strength is, is not
definitely solved and that much work is still needed before reaching the
final conclusion.

\end{document}